\documentclass[12pt]{article}
\usepackage{amsthm,amsfonts,amsmath,amssymb,mathrsfs,tikz,url,hyperref}

\title{Physical Meaning of Neumann and Robin Boundary Conditions for the Schr\"odinger Equation}

\author{
	Roderich Tumulka\footnote{Fachbereich Mathematik, Eberhard-Karls-Universit\"at T\"ubingen, Auf der Morgenstelle 10, 72076 T\"ubingen, Germany. E-mail: roderich.tumulka@uni-tuebingen.de}
}
\date{November 29, 2024}

\newcommand{\be}{\begin{equation}}
	\newcommand{\ee}{\end{equation}}

\newcommand{\CCC}{\mathbb{C}}

\newcommand{\RRR}{\mathbb{R}}

\newcommand{\ZZZ}{\mathbb{Z}}

\begin{document}
\maketitle
\begin{abstract}
The non-relativistic Schr\"odinger equation on a domain $\Omega\subset \RRR^d$ with boundary is often considered with homogeneous Dirichlet boundary conditions ($\psi(x)=0$ for $x$ on the boundary), homogeneous Neumann boundary conditions ($\partial_n \psi(x)=0$ for $x$ on the boundary and $\partial_n$ the normal derivative), or Robin boundary conditions ($\partial_n\psi(x)=\alpha\psi(x)$ for $x$ on the boundary and $\alpha$ a real parameter). Physically, the Dirichlet condition applies if the potential is much higher outside than inside the domain (``potential well''). We ask, when does the Neumann or Robin condition apply physically? Our answer is, when the potential is much lower (at the appropriate level) in a thin layer along the surface of a potential well, or when a negative delta potential of the appropriate strength is added at a surface close to the surface of the potential well.
		
\medskip
		
\end{abstract}

\section{Introduction}
\label{sec:intro}

We consider the non-relativistic one-particle Schr\"odinger equation
\be
i\hbar\frac{\partial \psi}{\partial t} = -\frac{\hbar^2}{2m} \nabla^2\psi +V\psi
\ee
on a domain $\Omega\subset \RRR^d$ with boundary in $d$ space dimensions, together with a boundary condition given by either the homogeneous Neumann condition
\be\label{N}
\partial_n\psi(x)=0
\ee
for $x$ on the boundary and $\partial_n$ the derivative in the direction orthogonal to the boundary, the Robin boundary condition
\be\label{R}
\partial_n\psi(x)=\alpha\psi(x)
\ee
with real constant $\alpha$ for $x$ on the boundary, or the homogeneous Dirichlet boundary condition
\be\label{D}
\psi(x)=0
\ee
for $x$ on the boundary. We ask in which physical situation each of these boundary conditions applies. For comparison, for the heat equation, the Neumann boundary condition is known to apply when the boundary surface is thermally insulated, Robin when the surface is in contact with a heat reservoir at temperature 0 through a wall with heat conductivity $\alpha$, and Dirichlet when the surface is in perfect thermal contact with a heat reservoir at temperature 0 \cite{wiki1}. 

The answer to our question is well known in the Dirichlet case: If we assume $\Omega$ to be in an infinite potential well, i.e., if we set $V(x)=\infty$ for $x\notin \Omega$, then $\psi$ vanishes outside $\Omega$, including on the boundary surface, and Eq.~\eqref{D} holds. This applies in practice when we surround $\Omega$ by walls made of a solid material while there is an (approximate) vacuum inside $\Omega$. There remains the natural and non-trivial question (raised, e.g., in \cite{Neal}) under which physical circumstances the other two boundary conditions \eqref{N}, \eqref{R} apply.

While a physical wave function is always defined everywhere in space $\RRR^d$, boundary conditions allow us to mathematically define the time evolution in the Hilbert space $L^2(\Omega,\CCC)$ instead of $L^2(\RRR^d,\CCC)$, which means that the wave functions will not even be defined outside $\Omega$. Mathematicians express this by saying that the Schr\"odinger Hamiltonian $-\tfrac{\hbar^2}{2m}\nabla^2+V$ together with any one of the boundary conditions \eqref{N}, \eqref{R}, \eqref{D} defines a self-adjoint operator $H$ in $L^2(\Omega,\CCC)$ \cite{Te}, and that any self-adjoint operator $H$ defines a unitary time evolution operator $U_t=\exp(-iHt/\hbar)$ \cite{sa}. The fact that the Neumann, Robin, and Dirichlet boundary conditions each define a unitary time evolution in $L^2(\Omega,\CCC)$ is why they are used in mathematical models (e.g., \cite{Te,BS23}).

Our claim here is that the Neumann condition \eqref{N} and the Robin condition \eqref{R} can be obtained by introducing a negative delta potential or a thin layer of huge negative potential just before a Dirichlet wall, as shown in Figure~\ref{fig:1} for $\Omega$ the negative half axis in 1 space dimension, and as expressed by means of formulas in Eq.s \eqref{V1} and \eqref{V2} below.


\begin{figure}[h]
	\begin{center}
	\begin{tikzpicture}
	\node at (-3,3) {a)};
	\draw[->] (-3,0) -- (1,0);
	\node at (0.8,-0.2) {$x$};
	\draw[->] (0,-3) -- (0,3);
	\node at (-0.6,2.7) {$V(x)$}; 
	\draw (-0.5,0)--(-0.5,0.2);
	\node at (-0.5,0.4) {$-L$}; 
	\draw[ultra thick] (-2.7,0) -- plot[domain=-3:3, samples=90] (-0.5+\x/20, {-3*exp(-\x*\x)}) -- (0,0)--(0,2.5);
	\end{tikzpicture}
\hspace{10mm}
	\begin{tikzpicture}
	\node at (-3,3) {b)};
	\draw[->] (-3,0) -- (1,0);
	\node at (0.8,-0.2) {$x$};
	\draw[->] (0,-3) -- (0,3);
	\node at (-0.6,2.7) {$V(x)$};
	\draw (-0.5,0)--(-0.5,0.2);
	\node at (-0.5,0.4) {$-L$}; 
	\draw[ultra thick] (-2.7,0)--(-0.5,0)--(-0.5,-2.5)--(0,-2.5)--(0,2.5);
	\end{tikzpicture}
	\end{center}
	\caption{a) Potential of Eq.~\eqref{V1}: Infinite well with additional term $\tfrac{\hbar^2}{2m}\lambda\,\delta(x+L)$ with small $L>0$ and large $\lambda<0$; the $\delta$ function has been drawn with finite width and height for better visibility. b) Potential of Eq.~\eqref{V2}: Infinite well with additional deep valley between $-L$ and 0. In both diagrams, $V(x)=\infty$ for $x> 0$, corresponding to a Dirichlet boundary condition at $x=0$.}
	\label{fig:1}
\end{figure}

The mathematical wave function, defined only in the region $\Omega$, is related to the physical wave function $\psi$, defined everywhere in space $\RRR^d$, as follows. To make the outside of $\Omega$ inaccessible to the particle, the strategy is to create a very high potential outside $\Omega$, or an infinite potential in an idealized description, which (as mentioned) leads to $\psi=0$ outside $\Omega$, and by continuity also $\psi=0$ on the boundary (the Dirichlet condition \eqref{D}). This might make it seem that Neumann and Robin conditions are physically impossible, but we show here how they are in fact possible in a certain limit, denoted in the following by $L\to 0$. For this demonstration, we start from the fact just explained that a Dirichlet condition \emph{can} physically be realized as the limit $V(x)\to\infty$ for $x\notin\Omega$.

We then argue, focusing on a prototypical example in $d=1$ space dimension, that the Hamiltonian $H_0$ on $\Omega=(-\infty,0]$ given by $-\tfrac{\hbar^2}{2m}\partial_x^2$ with Robin boundary condition \eqref{R} at $x=0$ \cite{convention}, which includes the Neumann condition for $\alpha=0$, can be obtained in the limit $L\to 0+$ (meaning $L\to 0$ while $L>0$) from the Hamiltonian $H_L$ on $\Omega=(-\infty,0]$ given by $-\tfrac{\hbar^2}{2m}\partial_x^2+V(x)$ with Dirichlet boundary condition at $x=0$ and potential either
\be\label{V1}
V=V_1(x) = -\tfrac{\hbar^2}{2m}\bigl(\tfrac{1}{L}+\alpha \bigr)\,\delta(x+L)
\ee
(see Figure~\ref{fig:1}a, $\delta$ is the Dirac delta function \cite{dimension}) or 
\be\label{V2}
V=V_2(x) = -\tfrac{\hbar^2}{2m} \bigl(\tfrac{\pi}{2L}+\tfrac{2\alpha}{\pi} \bigr)^2\; 1_{[-L,0]}(x)
\ee
(see Figure~\ref{fig:1}b, $1_A$ is the characteristic function of the set $A$). We conclude that the desired Hamiltonian $H_0$ can be physically realized by arranging the potential of Eq.~\eqref{V1} or \eqref{V2} with small $L>0$. 

\bigskip

\noindent{\bf Remark 1.} There is also an intuitive way of understanding why a potential valley in front of a Dirichlet wall can be relevant to producing the same effect as a Neumann condition. Since the Dirichlet condition forces $\psi$ to vanish on the boundary, it will effectively ``push'' the $|\psi|^2$ distribution away from the boundary towards the interior of $\Omega$. The Neumann condition, in contrast, has no such effect. So, in order to counter the ``repulsive'' effect of the Dirichlet condition, it seems fitting to add a potential valley in front of the boundary, as such a valley will tend to attract more of the wave function to the neighborhood of the boundary.

\bigskip

\noindent{\bf Remark 2.} It may seem puzzling that while physical wave functions are continuous, a wave function $\psi$ satisfying the Neumann or Robin conditions \eqref{N} or \eqref{R} will typically have non-zero values on the boundary of $\Omega$, so that, since we take $\psi=0$ outside $\Omega$, $\psi$ will have a jump discontinuity at the boundary. This aspect should be understood as follows. The physical wave function drops very rapidly to 0 as we cross the boundary in the outward direction, and it is a mathematical idealization to describe it as dropping infinitely rapidly, i.e., as having a jump discontinuity. In other words, while we show that the Neumann or Robin condition comes out in the limit $L\to 0$, the setup in reality will be close to but not exactly reach this limit. That is why there is no contradiction between continuous wave functions in reality and discontinuous ones in the mathematical treatment. (Similarly, we already used a mathematical idealization when we assumed the limit $V(x)\to \infty$ for $x\notin \Omega$.)

\bigskip

The remainder of this article is organized as follows. In Section~\ref{sec:eigenfunctions}, we compute the eigenvalues and eigenfunctions of the Hamiltonian, first using the Neumann or (more generally) Robin boundary condition, then $V_1$ as in Figure~\ref{fig:1}a, and then $V_2$ as in Figure~\ref{fig:1}b, thereby confirming the convergence claimed above. In Section~\ref{sec:lit}, we compare our results to those of previous work.
In Section~\ref{sec:conclusion}, we conclude.

\section{Eigenfunctions}
\label{sec:eigenfunctions}

In this section, we show that for $V$ as in Eq.~\eqref{V1} or \eqref{V2}, $H_L$ converges to $H_0$ as $L\to 0+$ in the sense that the eigenfunctions and eigenvalues of $H_L$ converge to those of $H_0$.

\subsection{With Robin Condition}

We now determine the eigenfunctions and eigenvalues of $H_0$ in order to later compare them to those of $H_L$. That is, here we are scrutinizing the Hamiltonian that we are going to justify later. Recall that for $H_0$, wave functions $\psi$ are defined only on $\Omega=(-\infty,0]$.

\subsubsection{Positive Eigenvalues}

The eigenfunctions of $-\tfrac{\hbar^2}{2m}\partial_x^2$ with positive eigenvalue $E>0$ are of the form
\be\label{eig}
\psi_k(x) = a_k\, e^{ikx} + b_k \,e^{-ikx}
\ee
for $k>0$ with eigenvalue $E=\tfrac{\hbar^2}{2m}k^2$ and complex coefficients $a_k,b_k$. Since we can change $\psi$ by an arbitrary global phase phase factor, we can and will take $a_k$ to be real. The function $\psi$ is not normalizable, which is common and unproblematical for eigenfunctions in the continuous spectrum (i.e., when there is a continuum of possible values for $k$ and thus for $E$). Since we do not have to normalize $\psi$, we
can and will choose $a_k=1$. The Robin condition \eqref{R} at $x=0$ is then equivalent to
\be\label{Rbk}
b_k = \frac{k+i\alpha}{k-i\alpha}
\ee
(the denominator is non-zero because $k\neq 0$), so the functions given by Eq.~\eqref{eig} with these coefficients are eigenfunctions of $H_0$. 

\subsubsection{Negative Eigenvalues}

For $\alpha\leq 0$, the above are already all of the eigenfunctions, but for $\alpha>0$ there exists one more, a normalizable bound state with energy $E<0$: Note first that the solutions of the eigenvalue equation $\-\tfrac{\hbar^2}{2m}\partial_x^2 \psi = E\psi$ for negative $E$ are of the form
\be\label{eigr}
\psi_r(x) = a_r \, e^{rx} + b_r \, e^{-rx}
\ee
with $r>0$ and $E=-\tfrac{\hbar^2}{2m}r^2<0$. Note next that $e^{rx}$ is normalizable on $(-\infty,0]$ (as $\int_{-\infty}^0 e^{2rx}dx= 1/2r$) but $e^{-rx}$ is not (as $\int_{-\infty}^0 e^{-2rx}dx=\infty$); as a consequence, $\psi_r$ is normalizable if and only if $b_r=0$ and normalized if in addition $a_r=\sqrt{2r}$, which we assume henceforth. The Robin condition \eqref{R} at $x=0$ then demands that $\sqrt{2r}re^{r0}=\alpha \sqrt{2r} e^{r0}$ or $r=\alpha$, which shows that there is only one possible $r$ value for $\alpha>0$ and none for $\alpha\leq 0$ since $r$ was supposed to be positive. For $\alpha>0$, the negative eigenvalue and the corresponding eigenfunction are
\be\label{eig0-}
E=-\tfrac{\hbar^2}{2m}\alpha^2~~\text{and}~~\psi(x)=\sqrt{2\alpha}\,e^{\alpha r}\,.
\ee
(For eigenvalue $E=0$, the eigenfunction has to be as in Eq.~\eqref{eig} with $k=0$ or, equivalently, Eq.~\eqref{eigr} with $r=0$, which is non-normalizable and therefore part of the continuous spectrum. However, this eigenfunction can be ignored since a single $k$ value in the continuous spectrum does not affect the operator in Hilbert space.)

\subsection{With delta Potential}

Let us turn to $V_1$. A delta potential $\tfrac{\hbar^2}{2m}\lambda\, \delta(x+L)$ with real prefactor $\lambda$ amounts to the boundary condition \cite{delta,BR14,Tum22}
\be\label{deltacond}
\partial_x\psi(-L+)-\partial_x\psi(-L-)= \lambda\psi(-L)\,,
\ee
where $\partial_x\psi(a\pm)$ means the right (left) derivative of $\psi$ at $a$, and $\psi$ is assumed continuous at $-L$. 

\subsubsection{Positive Eigenvalues}

Since away from $-L$, an eigenfunction of $H_L$ is just an eigenfunction of the Laplacian, we have for $E>0$ that
\be\label{eigenfunctionV1+}
\psi(x) = \begin{cases}
f_\mathrm{I}(x):= e^{ikx}+b_ke^{-ikx} & \text{for }x<-L\\
f_\mathrm{II}(x):= c_k e^{ikx} + d_k e^{-ikx} &\text{for }-L<x<0\,,
\end{cases}
\ee
where the wave number $k>0$ has to be the same in both regions to suit the same eigenvalue $E=\tfrac{\hbar^2}{2m}k^2>0$. The Dirichlet condition at $x=0$ enforces that $d_k=-c_k$, while the matching conditions at $x=-L$,
\begin{align}
f_\mathrm{I}(-L)&=f_\mathrm{II}(-L) \label{match1}\\
\partial_x f_\mathrm{II}(-L)-\partial_x f_\mathrm{I}(-L)
& =  \lambda\,f_\mathrm{II}(-L)\,,\label{match2}
\end{align}
then amount to
\begin{align}
e^{-ikL}+b_ke^{ikL}&=c_k(e^{-ikL}- e^{ikL})\\
c_k(ike^{-ikL} + ike^{ikL}) - (ike^{-ikL}-ikb_ke^{ikL}) &= \lambda c_k(e^{-ikL} - e^{ikL})\,,
\end{align}
which has exactly one solution given by
%
%
%
%
\begin{align}
b_k&= -e^{-2ikL} \frac{ke^{ikL}+\lambda \sin(kL)}{ke^{-ikL}+\lambda \sin(kL)}\label{bkdelta}\\
c_k&=  \frac{ke^{-ikL}}{ke^{-ikL}+\lambda \sin(kL)}\,.
\end{align}
(The denominator is non-zero because either its imaginary part $-k\sin(kL)$ is non-zero or else, whenever $kL$ is an integer multiple of $\pi$, the real part is $k\cos(kL)$, which is non-zero.)

Now we set $\lambda=-\bigl(\tfrac{1}{L}+\alpha \bigr)$ and then take the limit $L\to0$ (so $\lambda \to -\infty$); the interval $[-L,0]$ where $c_k$ applies shrinks to length 0, and for any fixed $k>0$,
\be
\exp(ikL)=1+ikL+\mathcal{O}(L^2)
\ee
(where $\mathcal{O}(L^n)$ means a term whose absolute value is $\leq C L^n$ for some constant $C$ and all sufficiently small $L$) and
\begin{align}
\lambda \sin(kL)&=-\bigl(\tfrac{1}{L}+\alpha \bigr)\bigl(kL+\mathcal{O}(L^3)\bigr)\\
&=-k-\alpha k L +\mathcal{O}(L^2) \,,
\end{align}
so
\begin{align}
b_k &\stackrel{\eqref{bkdelta}}{=} -\Bigl(1-2ikL+\mathcal{O}(L^2)\Bigr)\frac{k+ik^2L+\mathcal{O}(L^2) -k-\alpha kL + \mathcal{O}(L^2)}{k-ik^2L + \mathcal{O}(L^2)-k-\alpha kL + \mathcal{O}(L^2)}\\
&=-\underbrace{\Bigl(1-2ikL+\mathcal{O}(L^2)\Bigr)}_{\to 1} \frac{ik-\alpha + \mathcal{O}(L)}{-ik -\alpha + \mathcal{O}(L)}\\
&\to \frac{k+i\alpha}{k-i\alpha}
\end{align}
in agreement with Eq.~\eqref{Rbk}.

(As before, the case $E=0$ can be ignored because the eigenfunction, given by Eq.~\eqref{eigenfunctionV1+} with $k=0$, is non-normalizable, so zero dimensions of the Hilbert space correspond to $E=0$.)

\subsubsection{Negative Eigenvalues}

For $E<0$, eigenfunctions must be of the form
\be\label{eigenfunctionV1-}
\psi(x) = \begin{cases}
f_\mathrm{I}(x):= a_re^{rx}+b_r e^{-rx} & \text{for }x<-L\\
f_\mathrm{II}(x):= c_r e^{rx} + d_r e^{-rx} &\text{for }-L<x<0
\end{cases}
\ee
with $r>0$ and $E=-\tfrac{\hbar^2}{2m}r^2$. As before, the eigenfunction will be a bound state and as such must be normalizable, so $b_r=0$, and the Dirichlet condition at $x=0$ leads to $d_r=-c_r$. The matching conditions \eqref{match1}, \eqref{match2} now amount to
\begin{align}
a_r e^{-rL}&=c_r(e^{-rL}-e^{rL})\\
c_r(re^{-rL}+re^{rL})-a_rre^{-rL}&=\lambda a_re^{-rL}\,.
\end{align}
Solving for $c_r$ leads to
\begin{align}
c_r&= \frac{1}{1-e^{2rL}} a_r\\
c_r&= \frac{1+\lambda/r}{1+e^{2rL}} a_r
\end{align}
(where the denominators are clearly non-zero).
Since $a_r\neq 0$, the two equations can both hold only if 
\be\label{rcond1a}
\frac{1}{1-e^{2rL}} 
= \frac{1+\lambda/r}{1+e^{2rL}}\,,
\ee
a condition that restricts the possible values of $r$. 

Now set $\lambda=-(\tfrac{1}{L}+\alpha)$. Then condition \eqref{rcond1a} becomes, after some re-arrangement using that $\coth x=(e^x+e^{-x})/(e^x-e^{-x})=(e^{2x}+1)/(e^{2x}-1)$,
\be\label{rcond1b}
\coth(rL)-\frac{1}{rL} = -1 + \frac{\alpha}{r}\,.
\ee
Since, for any fixed $L>0$, the function $g(r):= \coth(rL) -\frac{1}{rL}$ for $r>0$ is continuous, positive, and strictly increasing and tends to 0 as $r\to0$ and to 1 as $r\to\infty$, its graph will intersect, for any fixed $\alpha>0$, the graph of $h(r):=-1+\frac{\alpha}{r}$ (which is continuous and strictly decreasing and tends to $\infty$ as $r\to0$ and to $-1$ as $r\to\infty$) exactly once, see Figure~\ref{fig:3}. For $\alpha\leq 0$, however, the graphs will not intersect because then $-1+\frac{\alpha}{r}\leq -1$ for all $r>0$. Thus, our Hamiltonian $H_L$ with $V_1$ has no negative eigenvalue for $\alpha\leq 0$ and exactly one for $\alpha>0$.

\begin{figure}[h]
	\begin{center}
	\begin{tikzpicture}
	\draw[->] (-0.5,0) -- (3,0);
	\node at (2.8,-0.2) {$r$};
	\draw[->] (0,-1.5) -- (0,2);
	\draw (-0.2,1)--(0,1);
	\node at (-0.5,1) {$1$};
	\draw[dashed] (0,1) -- (3,1); 
	\draw (-0.2,-1)--(0,-1);
	\node at (-0.5,-1) {$-1$};
	\draw[dashed] (0,-1) -- (3,-1); 
	\draw (1,-0.2) -- (1,0);
	\node at (1,-0.5) {$\alpha$};
	\draw[ultra thick] (0,0) -- plot[domain=0.1:2.99, samples=90] (\x, {1/tanh(1.5*\x)-1/(1.5*\x)});
	\node at (2.5,0.5) {$g$};
	\draw[ultra thick] plot[domain=0.34:2.99, samples=90] (\x, {-1+1/\x});
	\node at (0.6,1.7) {$h$};
	\end{tikzpicture}
	\end{center}
	\caption{The graphs of the functions $g(r)= \coth(rL) -\frac{1}{rL}$ and $h(r)=-1+\frac{\alpha}{r}$ for some choices of $L>0$ and $\alpha>0$; their intersection corresponds to the solution of Eq.~\eqref{rcond1b}.}
	\label{fig:3}
\end{figure}

We now determine the $r$ value satisfying Eq.~\eqref{rcond1b} in the limit $L\to 0+$. Since $\coth x -\frac{1}{x} \to 0$ as $x\to 0$, we have that $\coth(rL)-\frac{1}{rL} \to 0$ as $L\to 0$, and Eq.~\eqref{rcond1b} becomes $0=-1+\tfrac{\alpha}{r}$ or $r=\alpha$. Thus, the negative eigenvalue converges to $E=-\tfrac{\hbar^2}{2m}\alpha^2$, the interval $[-L,0]$ where $f_{\mathrm{II}}$ applies shrinks to length 0, $f_\mathrm{I}$ in Eq.~\eqref{eigenfunctionV1-} converges to $a_r e^{\alpha x}$, and the normalizing constant $a_r$ to $\sqrt{2\alpha}$, in agreement with Eq.~\eqref{eig0-}.

\subsection{With Potential Valley}
\label{sec:eigV2}

We now treat $V_2=-\tfrac{\hbar^2}{2m}v1_{[-L,0]}$ with prefactor $v>0$. 

\subsubsection{Positive Eigenvalues}

For $E>0$, an eigenfunction $\psi$ must be of the form 
\be
\psi(x) = \begin{cases}
f_\mathrm{I}(x):= e^{ikx}+b_ke^{-ikx} & \text{for }x<-L\\
f_\mathrm{II}(x):= c_k e^{iKx} + d_k e^{-iKx} &\text{for }-L<x<0
\end{cases}
\ee
with $K=\sqrt{k^2+v}$. The Dirichlet condition enforces again $d_k=-c_k$, and the matching conditions at $x=-L$ now read
\begin{align}
f_\mathrm{I}(-L)&=f_\mathrm{II}(-L)\label{match3}\\
\partial_x f_\mathrm{I}(-L)&=\partial_x f_\mathrm{II}(-L)\label{match4}
\end{align}
or
\begin{align}
e^{-ikL}+b_ke^{ikL}&=c_k (e^{-iKL} - e^{iKL})\\
ike^{-ikL}-ikb_ke^{ikL}&=c_k (iKe^{-iKL} +iK e^{iKL})\,,
\end{align}
which has the unique solution
\begin{align}
b_k&=\frac{k-iK\cot(KL)}{k+iK\cot(KL)} e^{-2ikL} \label{bkvalley}\\
c_k&=\frac{k}{K\cos(KL)-ik\sin(KL)}e^{-ikL}\,.
\end{align}
(The first denominator is non-zero because $k\neq 0$, the second because $\cos(kL)$ and $\sin(kL)$ are not simultaneously 0, while $k\neq 0$ and $K\neq 0$.)

Now we set
\be\label{vdef}
v=\Bigl(\frac{\pi}{2L}+\frac{2\alpha}{\pi} \Bigr)^2
\ee
and then take the limit $L\to0$; it follows that $v= \tfrac{\pi^2}{4L^2}+\tfrac{2\alpha}{L}+\mathcal{O}(L^0)$ and, since $k$ is fixed, $K=\tfrac{\pi}{2L}+ \tfrac{2\alpha}{\pi}+ \mathcal{O}(L)$, so $KL=\tfrac{\pi}{2}+\tfrac{2\alpha}{\pi}L+\mathcal{O}(L^2)$. Since $\cot(\tfrac{\pi}{2}+x)=-x+\mathcal{O}(x^3)$ for small $|x|$, we have that
\begin{align}
K\cot(KL)&= \bigl(\tfrac{\pi}{2L}+ \tfrac{2\alpha}{\pi}+ \mathcal{O}(L)\bigr)\bigl(-\tfrac{2\alpha}{\pi}L + \mathcal{O}(L^2)\bigr)\\
&= -\alpha+\mathcal{O}(L)
\end{align}
and thus
\begin{align}
b_k 
&\stackrel{\eqref{bkvalley}}{=} \frac{k+i\alpha+\mathcal{O}(L)}{k-i\alpha+\mathcal{O}(L)} \bigl(1+\mathcal{O}(L)\bigr)\\
&\to \frac{k+i\alpha}{k-i\alpha}
\end{align}
in agreement with Eq.~\eqref{Rbk}.

\subsubsection{Negative Eigenvalues}

For $E<0$, the treatment is more cumbersome; we give the details for the sake of completeness. For sufficiently large $v$, eigenfunctions must be of the form
\be\label{eigenfunctionV2-}
\psi(x) = \begin{cases}
f_\mathrm{I}(x):= a_re^{rx}+b_r e^{-rx} & \text{for }x<-L\\
f_\mathrm{II}(x):= c_r e^{iKx} + d_r e^{-iKx} &\text{for }-L<x<0
\end{cases}
\ee
with $r>0$, $K=\sqrt{v-r^2}$, and $E=-\tfrac{\hbar^2}{2m}r^2=\tfrac{\hbar^2}{2m}K^2-\tfrac{\hbar^2}{2m}v$. (For $r^2$ greater than $v$, $f_\mathrm{II}$ would contain $e^{Rx}$ with $R=\sqrt{r^2-v}$ instead of $e^{iKx}$, but this case is not relevant for us since we let $v\to\infty$.) As before, $b_r=0$, and the Dirichlet condition at $x=0$ leads to $d_r=-c_r$. The matching conditions \eqref{match3}, \eqref{match4} become
\begin{align}
a_re^{-rL}&=c_r (e^{-iKL} - e^{iKL})\label{match5}\\
a_rre^{-rL}&=c_r (iKe^{-iKL} +iK e^{iKL})\,,\label{match6}
\end{align}
and solving for $c_r$ leads to
\begin{align}
c_r&=\frac{ie^{-rL}}{2\sin(KL)}a_r\\
c_r&=\frac{-ire^{-rL}}{2K\cos(KL)}a_r\,.
\end{align}
(The denominators are never 0 because if $\sin(KL)=0$ then Eq.~\eqref{match5} implies that $a_r=0$ and Eq.~\eqref{match6} that $c_r=0$, and if $\cos(KL)=0$ then Eq.~\eqref{match6} implies that $a_r=0$ and Eq.~\eqref{match5} that $c_r=0$.)

As a consequence,
\be
\frac{ie^{-rL}}{2\sin(KL)}=\frac{-ire^{-rL}}{2K\cos(KL)}
\ee
or
\be\label{KcotKLr}
K\cot(KL)=-r\,,
\ee
which is a nonlinear equation in $r$ because $K=K(r)=\sqrt{v-r^2}$. (Recall that we consider only $r<\sqrt{v}$.)

Now set $v=(\frac{\pi}{2L}+\frac{2\alpha}{\pi})^2$ as in Eq.~\eqref{vdef}.

\bigskip

\noindent{\bf Claim.} {\it Suppose $0<L<\pi^2/4|\alpha|$. For $\alpha>0$,  Eq.~\eqref{KcotKLr} has a unique solution $0<r_0<\sqrt{v}$, while for $\alpha \leq 0$, it has no solution in this interval. The solution satisfies $\tfrac{\pi}{2L}<K(r_0)$.}

\begin{figure}[h]
	\begin{center}
	\begin{tikzpicture}
	\draw[->] (-0.5,0) -- (3.8,0);
	\node at (3.6,-0.3) {$K$};
	\draw[->] (0,-2.7) -- (0,1.5);
	\draw (-0.2,1) -- (0,1);
	\node at (-0.5,1) {$\tfrac{1}{L}$};
	\draw (3.1416,0)--(3.1416,0.2);
	\node at (3.1416,0.5) {$\tfrac{\pi}{L}$};
	\draw[dashed] (3.1416,0) -- (3.1416,-2.7); 
	\draw (1.5708,-0.2) -- (1.5708,0);
	\node at (1.5,-0.45) {$\tfrac{\pi}{2L}$};
	\draw[ultra thick] (0,1) -- plot[domain=0.1:2.4, samples=90] (\x, {\x*cot(\x r)});
	\node at (0.7,1.1) {$g$};
	\draw (1.9,0.2) -- (1.9,0);
	\node at (1.9,0.5) {$\sqrt{v}$};
	\draw[ultra thick] plot[domain=0:90, samples=46] ({1.9*sin(\x)},{-1.9*cos(\x)});
	\node at (0.7,-1.4) {$h$};
	\end{tikzpicture}
	\end{center}
	\caption{The graphs of the functions $g(K)= K\cot(KL)$ and $h(K)= -\sqrt{v-K^2}$ (a circular arc) for $v=(\tfrac{\pi}{2L}+\tfrac{2\alpha}{\pi})^2$ and some choices of $L>0$ and $\alpha>0$; their intersection corresponds to the solution of Eq.~\eqref{KcotKLr}.}
	\label{fig:4}
\end{figure}

\bigskip

\noindent{\it Proof.} For $L$ in the allowed range, $0<\frac{\pi}{2L}+\frac{2\alpha}{\pi}<\frac{\pi}{L}$ (in particular, $\sqrt{v}=\frac{\pi}{2L}+\frac{2\alpha}{\pi}$), and since $0<K<\sqrt{v}$, we have that $0<KL<\pi$, so $\cot(KL)$ is well defined. It is easiest to think of Eq.~\eqref{KcotKLr} as an equation in the variable $K$ by writing $r=\sqrt{v-K^2}$. The function $g(K):= K\cot(KL)$ is continuous and strictly decreasing on the interval $(0,\tfrac{\pi}{L})$, has limits $\frac{1}{L}$ as $K\to0$ and $-\infty$ as $K\to \frac{\pi}{L}$, and vanishes at $K=\frac{\pi}{2L}$ and only there, see Figure~\ref{fig:4}. The function $h(K):= -\sqrt{v-K^2}$ is continuous and strictly increasing on the interval $(0,\sqrt{v})$ and has limits $-\sqrt{v}$ as $K\to 0$ and $0$ as $K\to \sqrt{v}=\frac{\pi}{2L}+\frac{2\alpha}{\pi}$. For $\alpha\leq0$, $g>0$ and $h<0$ on the whole interval $(0,\sqrt{v})$ on which $h$ is defined, so their graphs do not intersect. For $\alpha>0$, it follows that they intersect exactly once (see Figure~\ref{fig:4}), say at $K_0$, which must be $>\tfrac{\pi}{2L}$ because $g\geq0$ for $K\leq \tfrac{\pi}{2L}$. The intersection point corresponds to $r_0=\sqrt{v-K^2_0}$.\hfill$\square$

\bigskip

We now compute for $\alpha>0$ the asymptotics of the solution $r_0$ as $L\to0$. Since $\tfrac{\pi}{2}<K_0L<\tfrac{\pi}{2}+\tfrac{2\alpha}{\pi}L$, it will be convenient to write $K_0L= \tfrac{\pi}{2}+\theta L$ with $0<\theta<\tfrac{2\alpha}{\pi}$. Since $\cot(\tfrac{\pi}{2}+x)=\mathcal{O}(x)$ and $K_0L=\tfrac{\pi}{2}+\mathcal{O}(L)$, we have that $K_0L\cot(K_0L)=\mathcal{O}(L)$. By Eq.~\eqref{KcotKLr}, $K_0L\cot(K_0L) = -\sqrt{vL^2-K_0^2L^2}$, so
\begin{align}
\mathcal{O}(L^2)
&=vL^2-K_0^2L^2\\
&\stackrel{\eqref{vdef}}{=} (\tfrac{\pi}{2} + \tfrac{2\alpha}{\pi} L)^2 - (\tfrac{\pi}{2}+\theta L)^2\\
&= 2\alpha L - \pi\theta L+\mathcal{O}(L^2)\,.
\end{align}
Dividing by $L$ and rearranging yields that
\be
\theta= \frac{2\alpha}{\pi} + \mathcal{O}(L)
\ee
or
\be
K_0L = \frac{\pi}{2} + \frac{2\alpha}{\pi}L + \mathcal{O}(L^2)\,.
\ee
In order to convert that into the asymptotics of $r_0$, it is easier to use Eq.~\eqref{KcotKLr} than $r_0=\sqrt{v-K_0^2}$. Since $\cot (\tfrac{\pi}{2}+x)=-x + \mathcal{O}(x^3)$ for small $|x|$, $\cot(K_0L)= -\tfrac{2\alpha}{\pi}L+\mathcal{O}(L^2)$ and thus
\be
r_0 = -K_0 \cot(K_0L) = \alpha+\mathcal{O}(L)\,,
\ee
so $r_0$ tends to $\alpha$ as $L\to 0$. Thus, the negative eigenvalue converges to $E=-\tfrac{\hbar^2}{2m}\alpha^2$ and the eigenfunction to $\sqrt{2\alpha} e^{\alpha x}$, in agreement with Eq.~\eqref{eig0-}.

\section{Notes on the Literature}
\label{sec:lit}

I am not aware of a treatment in the literature of the potential $V_1$ involving the delta function, but the potential $V_2$ or variants thereof have been considered in \cite{S85,FCT01,BW10}. Here are some remarks on these articles and their relation to the considerations above.

\v{S}eba \cite{S85} addressed a more general and more complicated problem (viz., the scaling limit $\varepsilon \to 0+$ for $-\tfrac{\hbar^2}{2m}\partial_x^2+\varepsilon^{-2}\lambda(\varepsilon)V(x/\varepsilon)$ on the half line), which however includes the convergence of our $H_L$ with $V_2=-\tfrac{\hbar^2}{2m}v1_{[-L,0]}$ as a special case. In fact, he mentioned this case explicitly, but obtained the Robin Hamiltonian $H_0$ only for discrete special values \cite{special}
\be\label{special}
\alpha=(2n+1)\frac{\pi}{2}\text{ with }n=0,1,2,\ldots\,.
\ee
(Although he did not say explicitly that other values could not be obtained, readers of the paper can at least easily get this impression.) Actually, he made an assumption ($\lambda(\varepsilon)=1+\varepsilon$ on page 23) that is more restrictive than necessary, and with a different assumption ($\lambda(\varepsilon)=1+(\mathrm{const.})\varepsilon$) he would have obtained the Robin Hamiltonian for arbitrary $\alpha\in\RRR\setminus\{0\}$. (The $V_2$ with $\alpha=0$, from which we obtain the Neumann Hamiltonian, is not covered by his results.)

Belchev and Walton \cite{BW10} also obtained the Robin Hamiltonian as a limit of a potential as in Figure~\ref{fig:1}b, but like \v{S}eba only for the discrete values given by Eq.~\eqref{special}. They explicitly claimed that the Robin Hamiltonian arises  only for these discrete values, but that is not correct, as we saw in Section~\ref{sec:eigenfunctions}. (In their Eq.~(10), they made the too restrictive assumption that $v$ is of the form (in their notation) $\kappa^2 \alpha\ell(\alpha\ell +1)$, while the expression $\kappa^2 \alpha\ell(\alpha\ell+\mathrm{const.})$ (or varying the parameter $\ell$ while taking the limit) would have led to arbitrary Robin conditions, including the Neumann condition.)

F\"ul\"op, Cheon, and Tsutsui \cite{FCT01} considered $-\tfrac{\hbar^2}{2m}\partial_x^2+V(x)$ on the whole real line with a potential $V$ similar to $V_2$ but with a large but finite constant $V_0>0$ for $x>0$, that is, $V(x)=-\tfrac{\hbar^2}{2m}v1_{[-L,0)}+ V_0 1_{[0,\infty)}$ (except that they wrote $-x$ for our $x$). They obtained the Robin Hamiltonian $H_0$ in a suitable limit involving $L\to 0$, $v\to \infty$, $V_0\to\infty$, but the calculation gets much more complicated than ours by considering finite $V_0$.

Broussard \cite{Br09} provided a physical example in which a function $f(x)$ obeys an equation equivalent to a Schr\"odinger equation on an interval $[a,b]$ with Neumann boundary conditions, but this function $f$ is not a quantum-mechanical wave function.

Another situation in which the Neumann boundary condition arises has been in the literature before \cite[Sec.~1.4]{AGHKH88}: There is a particular zero-range interaction called the ``$\delta'$ interaction,'' and such an interaction at the origin for the Hamiltonian $-\tfrac{\hbar^2}{2m}\partial_x^2$ on $\Omega=\RRR$ would correspond to the boundary condition
\be\label{delta'bc}
\partial_x\psi(0+)=\partial_x\psi(0-)=\beta\bigl(\psi(0+)-\psi(0-)\bigr)\,,
\ee
where $\beta$ is a real parameter and $\psi(0\pm)=\lim_{\varepsilon\to 0+}\psi(\pm\varepsilon)$ is the right (left) value at a jump discontinuity. For $\beta=0$, we obtain a Hamiltonian for which no part of $\psi$ can propagate from the left half axis to the right or vice versa, and each half is subject to a Neumann condition at the origin. However, while the $\delta$ potential corresponds to multiplication by a Dirac $\delta$ distribution, the ``$\delta'$ interaction,'' despite its name, does not actually correspond to a potential given by (a multiple of) the derivative of $\delta$. (Rather, $H=-\tfrac{\hbar^2}{2m}\partial_x^2 +\tfrac{\hbar^2}{2m\beta}\delta'(x)$ would lead, when applied to functions with jump discontinuities at the origin in the zeroth and first derivative, to divergent terms, and only after those are artificially removed, to the condition \eqref{delta'bc}.) As a consequence, it is not clear how it could be realized physically.

\section{Conclusion}
\label{sec:conclusion}

When the non-relativistic Schr\"odinger equation is considered on a domain $\Omega$ with a boundary, the time evolution of the wave function $\psi$ is determined only when a \emph{boundary condition} has been specified in addition to the demand that the Hamiltonian acts like $-\tfrac{\hbar^2}{2m}\nabla^2 + V$. Equivalently, the specification of a \emph{self-adjoint operator} includes the specification of a boundary condition. The most common boundary conditions leading to a self-adjoint Hamiltonian (and thus a unitary time evolution) are the homogeneous Dirichlet, Neumann, and Robin boundary conditions \cite{Te}. Ordinary reflecting walls correspond to a sudden large increase in the potential, idealized as $V=+\infty$ outside $\Omega$, which mathematically means a Dirichlet boundary condition ($\psi(x)=0$ for $x$ on the boundary). This leads to the question, which physical situation, or which kind of wall, corresponds to a Neumann or Robin condition? 

We have shown that a wall covered with a thin layer providing a strong negative potential, either as in Figure~\ref{fig:1}a or as in Figure~\ref{fig:1}b, with the appropriate relation (as in Eq.~\eqref{V1}, respectively \eqref{V2}) between the thickness $L$ and the strength $\lambda$ (respectively $v$), leads to the desired boundary condition. We did so by showing that the eigenfunctions and eigenvalues of the Hamiltonian with the appropriate potential, given by Eq.~\eqref{V1} or \eqref{V2}, approach, in the limit $L\to 0$ (while $\lambda\to -\infty$, respectively $v\to\infty$ at the appropriate rate), those of $-\tfrac{\hbar^2}{2m}\nabla^2$ with Neumann or Robin boundary condition. This result will also carry over to $(d-1)$-dimensional boundaries of $d$-dimensional regions $\Omega$ by using the potential of Eq.~\eqref{V1} or \eqref{V2} with $x$ the negative distance from the boundary (so that the direction in which $x$ increases is orthogonal to the boundary).


\begin{thebibliography}{12}
\bibitem{dimension} The values of the delta function have the dimension 1/length; since $\alpha$ also has the dimension 1/length, $V$ has the dimension of an energy, as it should.
\bibitem{delta} The condition \eqref{deltacond} can be understood as follows. If $\psi$ is an eigenfunction, so $(-\partial_x^2+\lambda\, \delta(x+L))\psi(x) = \tfrac{2m}{\hbar^2}E\psi(x)$, then integrating both sides from $L-\varepsilon$ to $L+\varepsilon$ and letting $\varepsilon\to 0+$ leads to Eq.~\eqref{deltacond}.
\bibitem{special} The values in Eq.~\eqref{special} are the positive zeroes of the cotangent function. In Section~\ref{sec:eigV2}, it played a role that $KL$ converges to a zero of the cotangent, and we chose the level $v$ of the potential valley in $V_2$ in such a way that $KL\to\tfrac{\pi}{2}$, but we could have used a different zero at $(2n+1)\tfrac{\pi}{2}$ with $n\in\ZZZ$. This would come up for $v=\bigl(\tfrac{(2n+1)\pi}{2L}+\tfrac{2\alpha}{(2n+1)\pi} \bigr)^2$ with $n\in\{0,1,2,\ldots\}$ instead of the $v$ of Eq.~\eqref{vdef}. For this choice with $n\neq 0$, most of the calculation of Section~\ref{sec:eigV2} goes through in the same way, but Eq.~\eqref{KcotKLr} has further solutions $r$ which tend to $\infty$ and whose eigenfunctions do not converge in Hilbert space. Then $H_L$ does not converge to $H_0$ in the sense of convergence of the eigenvalues and eigenfunctions, but it does in other senses of convergence of operators, such as the norm resolvent sense used by \v{S}eba \cite{S85}.
\end{thebibliography}
\end{document}